\documentclass[runningheads]{llncs}
\usepackage[utf8]{inputenc}
\usepackage[T1]{fontenc}
\usepackage{lmodern}
\usepackage{xcolor}
\definecolor{bestgreen}{HTML}{1A5C2E}

\usepackage{amsmath,amssymb,mathtools}
\usepackage{bm} 

\usepackage{graphicx}
\usepackage{booktabs}    
\usepackage{multirow}
\usepackage{makecell}
\usepackage{array}
\usepackage{url}
\usepackage{hyperref}
\usepackage[nameinlink,noabbrev]{cleveref} 
\usepackage[numbers]{natbib}
\usepackage{algorithm}
\usepackage{algpseudocode}
\usepackage{adjustbox}
\usepackage{placeins}
\usepackage{tikz}
\usetikzlibrary{positioning,calc,arrows.meta,fit,backgrounds,shadows.blur,decorations.pathreplacing}
\definecolor{indigo}{RGB}{76,81,191}  

\makeatletter
\let\origthefootnote\thefootnote
\renewcommand\thefootnote{\fnsymbol{footnote}} 
\newcommand{\samethanks}[1][\value{footnote}]{\footnotemark[#1]} 
\makeatother

\newcommand{\heatcell}[1]{#1}
\begin{document}
\title{Contradictions in Context: Challenges for Retrieval-Augmented Generation in Healthcare}
%
%
%
%
%
\titlerunning{Toward Safer Medical RAG}

\author{%
  Saeedeh Javadi\inst{1,2}\thanks{Equal contribution.} \and
  Sara Mirabi\inst{1}\samethanks \and
  Manan Gangar\inst{1} \and
  Bahadorreza Ofoghi\inst{1}
}
 \authorrunning{Javadi et al.}

\institute{%
  Deakin University, 221 Burwood Highway, Burwood, Victoria 3125, Australia\\
  \and 
    RMIT University, 124 La Trobe Street, Melbourne, Victoria 3000, Australia\\
  \email{saeedeh.javadi@student.rmit.edu.au}\\
  \email{s.mirabi@deakin.edu.au, gangarmanan27@gmail.com, b.ofoghi@deakin.edu.au}
}

\maketitle             
\makeatletter
\let\thefootnote\origthefootnote
\setcounter{footnote}{0}
\makeatother
\pagestyle{plain}
%




\begin{abstract}
In high-stakes information domains such as healthcare, where large language models (LLMs) can produce hallucinations or misinformation, retrieval-augmented generation (RAG) has been proposed as a mitigation strategy, grounding model outputs in external, domain-specific documents. Yet, this approach can introduce errors when source documents contain outdated or contradictory information. This work investigates the performance of five LLMs in generating RAG-based responses to medicine-related queries. Our contributions are three-fold: i) the creation of a benchmark dataset using consumer medicine information documents from the Australian Therapeutic Goods Administration (TGA), where headings are repurposed as natural language questions, ii) the retrieval of PubMed abstracts using TGA headings, stratified across multiple publication years, to enable controlled temporal evaluation of outdated evidence, and iii) a comparative analysis of the frequency and impact of outdated or contradictory content on model-generated responses, assessing how LLMs integrate and reconcile temporally inconsistent information. Our findings show that contradictions between highly similar abstracts do, in fact, degrade performance, leading to inconsistencies and reduced factual accuracy in model answers. These results highlight that retrieval similarity alone is insufficient for reliable medical RAG and underscore the need for contradiction-aware filtering strategies to ensure trustworthy responses in high-stakes domains.

\keywords{Large language models \and Retrieval-augment generation \and Contradiction detection \and Medicine \and Question answering.}
\end{abstract}

\section{Introduction}
Large language models (LLMs) have demonstrated exceptional capabilities in responding to user information requests that require world knowledge. This knowledge is inherent in the vast amounts of pre-training data they consume, which is increasingly making these models a primary candidate for information seeking, where they have achieved state-of-the-art performance on a wide range of tasks, including medical question answering (QA)~\cite{Xiong0LZ24}. In domains like health, however, knowledge of specific concepts is constantly being updated with new findings, some of which are contradictory to previously known facts~\cite{carpenter2016,abcnews2018}. 
Inconsistent and outdated sources used for training the generative models can potentially result in unreliable generated information. To mitigate misinformation and hallucinations resulting from such inconsistencies in training medical information where accuracy is paramount~\cite{pal-etal-2023-med}, retrieval-augmented generation (RAG) is employed as a strategy~\cite{mkrag2025,Xiong0LZ24}. RAG effectively supplements the LLM's internal knowledge with context information, and studies have shown that augmenting prompts with retrieved evidence can reduce the incidence of hallucinations in practice~\cite{NEURIPS2020_6b493230}. The efficacy of RAG is, however, predicated on the relevance and coherence of the retrieved documents, thereby raising significant concerns regarding the model's performance and robustness should the retrieval process yield contradictory or irrelevant results~\cite{yan2024}. 
A significant challenge for RAG systems, therefore, is the LLM's handling of conflicting or outdated information within retrieved documents~\cite{XuQGW0ZX24}. Given that an LLM's internal parameters are static, conflicts between its fixed knowledge base and new, external context are an unavoidable issue. 
Resolving such conflicts is a non-trivial task, as the model may incorrectly prioritize a particular source or even synthesize contradictory information. Despite this, there is currently a lack of systematic research and established guidelines for managing these issues in RAG systems, particularly within medical applications~\cite{Xiong0LZ24}.

In this work, we investigate how well LLMs handle medicine-related queries when grounded in authoritative reference documents. We focus on the effects of contradictions and outdated content on the quality of RAG-powered LLM outputs using a purpose-curated dataset from two prominent sources: the Australian Therapeutic Goods Administration (TGA) and the PubMed repository. The dataset includes medicine-related questions, linked PubMed abstracts, and metadata such as publication year, ensuring a mix of recent and older sources. The questions were used as information requests to several LLMs, where the relevant PubMed abstracts were utilized for RAG, and the correct answers from the TGA brochures were considered as ground-truth responses. 
\footnote{The codes and datasets are available at the GitHub repository \url{https://github.com/pub2026/MedicalContradictionDetection-RAG}}

To make the study objectives explicit and address this gap, we frame our investigation around the following research questions:
\begin{itemize}
    \item \textbf{RQ1:} How do outdated or temporally inconsistent PubMed abstracts affect the accuracy and consistency of RAG-generated responses?
    \item \textbf{RQ2:} How do different retrieval conditions, specifically most similar, most contradictory, and least contradictory, influence the correctness of the generated answers?
    \item \textbf{RQ3:} Do different LLM architectures exhibit varying levels of robustness when synthesizing information from contradictory sources?
\end{itemize}

\vspace{5mm}
Our main contributions include:

\begin{itemize}
    \item Curating a dataset linking TGA consumer medicine information with PubMed abstracts across multiple publication years, prioritizing temporal diversity.
    \item Implementation of a RAG pipeline using FAISS retrieval and the BAAI/bge-small-en-v1.5 embedding model, enabling reproducible evaluation.
    \item Evaluation of five LLMs, including Falcon3~\cite{falcon3}, Gemma-3~\cite{gemma3-270m-it}, GPT-OSS~\cite{gpt-oss-20b}, Med-LLaMA3~\cite{med-llama3-8b}, and Mixtral~\cite{mixtral-8x7b} on the dataset to assess their ability to provide accurate, up-to-date, and non-contradictory answers using RAG.
    \item Analysis of how the time stamp of source information and potential contradictions in source materials affect the quality of generated responses.
\end{itemize}


\vspace{-6pt}
\section{Related Work}
Several studies have taken RAG strategies in the domain in recent years. MKRAG, a RAG framework for medical QA, retrieves medical facts from an external disease database and injects them into LLM prompts through in-context learning~\cite{shi2025mkrag}. OpenEvidence and ChatRWD were developed to apply RAG to literature-based clinical evidence and map queries into PICO study designs for real-world evidence generation~\cite{low2024answering}. In a similar direction, a RAG-driven model was proposed for health information retrieval that integrates PubMed Central with generative LLMs through a three-stage pipeline of passage retrieval, generative response construction, and factuality checks via stance detection and semantic similarity~\cite{upadhyay2025enhancing}. A multi-source benchmark for evidence-based clinical QA was curated from Cochrane reviews~\cite{cochrane2025library}, AHA guidelines~\cite{aha2025guidelines}, and narrative guidance, underscoring the need for diversified and authoritative retrieval sources~\cite{wang2025evaluating}. Extending this line, MedCoT-RAG combined causal-aware retrieval with structured chain-of-thought prompting for medical QA~\cite{wang2025medcot}. A two-layer RAG framework leveraging Reddit data on emerging substances such as xylazine and ketamine was introduced to generate query-focused summaries suitable for low-resource settings~\cite{das2025two}. BriefContext, a map-reduce framework that partitions long retrieval contexts into shorter segments to mitigate the ``lost-in-the-middle'' problem and improve the accuracy of medical QA in RAG systems, was introduced \cite{zhang2025leveraging}. EXPRAG, a retrieval-augmented generation framework, leveraged electronic health records by retrieving discharge reports from clinically similar patients through a coarse-to-fine process (EHR-based report ranking followed by experience retrieval), enabling case-based reasoning for diagnosis, medication, and discharge instruction QA~\cite{ou2025experience}. MedRAG integrated electronic health records with a hierarchical diagnostic knowledge graph for clinical decision support~\cite{zhao2025medrag}. Discuss-RAG, an agent-led framework, enhanced RAG by using multi-agent discussions to construct context-rich summaries for retrieval and a verification agent to filter irrelevant snippets before answer generation \cite{dong2025talk}. Finally, a biomedical QA system based on RAG was developed with a two-stage hybrid retrieval pipeline, where BM25~\cite{robertson2009probabilistic} performed lexical retrieval over PubMed, and MedCPT’s cross-encoder~\cite{jin2023medcpt} reranked the top candidates to refine semantic relevance~\cite{stuhlmann2025efficient}.

While previous studies demonstrate the benefits of RAG in medical QA, most concentrate on retrieval pipelines, evidence structuring, or database integration. Few explicitly examine the risks posed by outdated or contradictory evidence within retrieved sources. This remains a critical gap for high-stakes medical applications, where knowledge evolves rapidly. 

\vspace{-5pt}
\section{Methodology}\vspace{-3pt}
\subsection{Problem Formulation and Objectives}
This study addresses the challenge of evaluating RAG in the medical QA domain, where information accuracy and temporal consistency are critical. The problem is formally defined as follows.
Let \[
\mathcal{M}=\{m_1,\ldots,m_{1476}\},
\tag{1}
\] denote a set of $1{,}476$ medicines regulated by the TGA. For each medicine $m_i\in\mathcal{M}$, a fixed set of six standardized consumer-oriented queries, 
\[
\mathcal{Q}_i=\{\,q_{i,1},q_{i,2},\ldots,q_{i,6}\,\},
\tag{2}
\] is defined, which covers common information needs: 1) therapeutic indications, 2) pre-use warnings, 3) drug-drug interactions, 4) dosage and administration, 5) guidance while under treatment, including interactions/monitoring, and 6) adverse effects.
\footnote{The sixth standardized phrasings in our corpus include, e.g., \emph{``Why am I using \texttt{ABACAVIR}?''} (indications); \emph{``What should I know before I use \texttt{ABACAVIR}?''} (pre-use warnings); \emph{``What if I am taking other medicines with \texttt{ABACAVIR}?''} (drug-drug interactions); 
\emph{``How do I use \texttt{ABACAVIR}?''} (dosage/administration); \emph{``What should I know while using \texttt{ABACAVIR}?''} (on-treatment guidance); and \emph{``Are there any side effects of \texttt{ABACAVIR}?''} (adverse effects).} These queries represent typical information needs that patients and healthcare consumers might have when consulting medical information systems.
Across the collection, a total of
$|\mathcal{Q}|=\sum_{i=1}^{1476}|\mathcal{Q}_i|=6\,|\mathcal{M}|=8{,}856$ 
query instances are obtained. For each query $q_{i,j}\in\mathcal{Q}_i$ with $j\in\{1,\ldots,6\}$, a retrieval process is  performed against the PubMed biomedical literature database to obtain a candidate set of abstracts, \[
\mathcal{D}_{i,j}=\{\,d_1,\ldots,d_{n_{i,j}}\,\},
\tag{3}
\]
where each document $d\in\mathcal{D}_{i,j}$ is characterized by its text $t(d)$, publication year $\gamma(d)$, citation count $\kappa(d)\in\mathbb{N}_0$, and unique PubMed identifier $\mathrm{PMID}(d)$. The retrieval process yields resulting in a total corpus containing approximately $400{,}000$ documents spanning publication years from $1975$ to $2025$. The fundamental challenge lies in the fact that retrieved abstracts may contain outdated recommendations, conflicting findings from different studies, or evolving medical consensus over time. The RAG system must therefore not only identify relevant documents but also reconcile potentially contradictory information while generating accurate, consistent responses. Our primary objectives are therefore threefold.

\vspace{6pt}
\noindent\textbf{Objective 1: Temporal Diversity in Evidence Selection.} Given the evolving nature of medical knowledge, the system must select a subset, \[
\mathcal{R}_{i,j}\subseteq \mathcal{D}_{i,j}, \qquad |\mathcal{R}_{i,j}|\leq 20,
\tag{4}
\] that maximizes temporal diversity while maintaining relevance. This requires balancing recent findings with historical context, particularly when medical recommendations change over time.

\vspace{6pt}
\noindent\textbf{Objective 2: Contradiction-Aware Retrieval.} The system must identify and quantify contradictions among retrieved abstracts to avoid incorrect or potentially harmful responses. A contradiction function,
\[
\mathcal{CNT}:\mathcal{R}_{i,j}\times\mathcal{R}_{i,j}\to[0,1],
\tag{5}
\]
is defined such that, for any pair $(d_a,d_b)$, the value $\mathcal{CNT}(d_a,d_b)$ quantifies the likelihood that $t(d_a)$ contradicts $t(d_b)$; this function is intended to support the construction of ``most-contradictory’’ and ``least-contradictory’’ configurations.

\vspace{6pt}
\noindent\textbf{Objective 3: RAG Performance Evaluation.}
Given a retrieved set $\mathcal{R}_{i,j}$ and an LLM $L\in\Lambda$, the system must generate a response $a_{i,j}$ that addresses $q_{i,j}$ \emph{using the retrieved context}. The quality of this response is evaluated against ground truth answers extracted from TGA medicine information documents. Five LLMs including Falcon3~\cite{falcon3}, Gemma-3~\cite{gemma3-270m-it}, GPT-OSS~\cite{gpt-oss-20b}, Med-LLaMA3~\cite{med-llama3-8b}, and Mixtral~\cite{mixtral-8x7b} are evaluated, collectively denoted as $\Lambda$.

\vspace{-6pt}
\subsection{Evidence Set Construction}\label{sec:evidence}

\subsubsection{Query Expansion and Search.}
The evidence acquisition employs a three-tier query formulation strategy to balance precision and recall. For each $(m_i, q_{i,j})$ pair, content terms $\mathcal{T}_{i,j} = \{t_1, t_2, \ldots, t_k\}$ are extracted through part-of-speech filtering using SpaCy \cite{honnibal2020spacy}, retaining nouns (NN, NNS), verbs (VB*), and proper nouns (NNP, NNPS) while excluding pharmaceutical company names through a curated exclusion list. Three query formulations are constructed: i) exact sentence match $Q_1 = \bigwedge_{t \in \mathcal{T}_{i,j}} \text{sentence}(t)$, ii) proximity-constrained $Q_2 = \text{NEAR}_{25}\allowbreak(\mathcal{T}_{i,j}) \wedge m_i[\text{ti}]$, and iii) full query proximity $Q_3 = \text{NEAR}_{25}(q_{i,j})$,
where $\text{NEAR}_k$ denotes proximity within $k$ tokens and $[\text{ti}]$ restricts to title field. Results are deduplicated by PMID to form $\mathcal{D}_{i,j}^{\text{raw}}$.

\vspace{-12pt}
\subsubsection{Abstract Acquisition and Filtering.}
PMIDs are fetched in batches to mitigate API limitations. XML responses are parsed to extract $(\mathrm{pmid},\gamma(d),t(d))$. Records without abstracts are discarded; all remaining years are preserved to enable temporal analyses.

\vspace{-12pt}
\subsubsection{Temporal-Citation Balanced Selection.}
A critical innovation in this methodology is the implementation of a selection algorithm that balances temporal diversity with citation impact. This approach addresses the challenge of capturing evolving medical knowledge while prioritizing influential research within each temporal stratum. Based on Algorithm~\ref{alg:year-cite}, to construct a pool of up to 20 abstracts per query while preserving temporal coverage, when the number of unique publication years exceeds 20, the system performs stratified sampling to select years with approximately three-year intervals, ensuring representation across the full temporal range. For datasets with fewer than 20 unique years, all years are retained. Within each selected year, abstracts are ranked by citation count in descending order, serving as a proxy for scientific impact and reliability.

The final selection employs a round-robin approach across years, iteratively selecting the highest-cited unselected abstract from each year until either 20 abstracts are selected or all available abstracts are exhausted. This mechanism ensures both temporal diversity and quality, preventing over-representation of any single time period while maintaining citation-based quality signals.

\begin{algorithm}[]
\caption{Temporal-citation balanced selection}\label{alg:year-cite}
\begin{algorithmic}[1]
\Require Candidates $\mathcal{D}_{i,j}^{\text{raw}}$, citation function $\kappa : \mathcal{D} \to \mathbb{N}$, year function $\gamma : \mathcal{D} \to \mathbb{N}$
\Ensure Selected subset $\mathcal{R}_{i,j}$ where $|\mathcal{R}_{i,j}| \leq 20$
\State $\mathcal{Y} \gets \{\gamma(d) : d \in \mathcal{D}_{i,j}^{\text{raw}}\}$
\If{$|\mathcal{Y}| \geq 20$}
    \State $\mathcal{Y}^* \gets \text{STRATIFIED-SAMPLE}(\mathcal{Y}, 20, \text{gap}=3)$
\Else
    \State $\mathcal{Y}^* \gets \mathcal{Y}$
\EndIf
\ForAll{$y \in \mathcal{Y}^*$}
    \State Sort $\{d \in \mathcal{D}_{i,j}^{\text{raw}} : \gamma(d) = y\}$ by $\kappa(d)$ descending
\EndFor
\State $\mathcal{R}_{i,j} \gets \emptyset$
\While{$|\mathcal{R}_{i,j}| < 20$ and $\exists y \in \mathcal{Y}^*$ with remaining candidates}
    \ForAll{$y \in \mathcal{Y}^*$ in round-robin order}
        \If{candidates remain for year $y$}
            \State $\mathcal{R}_{i,j} \gets \mathcal{R}_{i,j} \cup \{\arg\max_{d:\gamma(d)=y} \kappa(d)\}$
        \EndIf
    \EndFor
\EndWhile
\State \Return $\mathcal{R}_{i,j}$ 
\end{algorithmic}
\end{algorithm}

\subsection{Diversity-Aware Scoring for Retrieval Framework}\label{sec:diversity}

\subsubsection{Document Representation and Indexing.}\label{sec:representation}

Document embeddings are computed using an encoder function $e: \Sigma^* \to \mathbb{R}^d$, specifically the BAAI/bge-small-en-v1.5 model \cite{xiao2023bge}. 
The embeddings are indexed using Facebook AI Similarity Search (FAISS) \cite{johnson2019billion}.

\subsubsection{Maximal Marginal Relevance with Temporal Augmentation.}
The retrieval framework extends maximal marginal relevance (MMR) \cite{carbonell1998use} by incorporating temporal diversity. For a given query $q$ and candidate set $\mathcal{D}$, the MMR score for document $d_i$ is:
\begin{equation}
\text{MMR}(d_i \mid q, \mathcal{D}) = \lambda \cdot \cos(e(q), e(t(d_i))) - (1-\lambda) \cdot \max_{j \neq i} \cos(e(t(d_i)), e(t(d_j)))
\tag{6}
\end{equation}
where cosine similarity is defined as
${\cos(\mathbf{x}, \mathbf{y}) = \frac{\mathbf{x}^\top \mathbf{y}}{\|\mathbf{x}\|_2 \|\mathbf{y}\|_2}}$, 
in the embedding space, and $\lambda$ controls the trade-off between relevance and diversity. This score is then combined with a temporal diversity component. A temporal diversity score is computed to favor documents spanning different time periods:

\begin{equation}
\tau(d_i) = \frac{\gamma(d_i) - \min_{d \in \mathcal{D}} \gamma(d)}{\max_{d \in \mathcal{D}} \gamma(d) - \min_{d \in \mathcal{D}} \gamma(d) + \epsilon}
\tag{7}
\label{eq:year}
\end{equation}

\noindent where $\epsilon = 10^{-5}$ prevents division by zero. The final ranking score integrates relevance, redundancy, and temporal components:

\begin{equation}
S(d_i \mid q) = \alpha \cdot \text{MMR}(d_i \mid q, \mathcal{D}) + (1-\alpha) \cdot \tau(d_i)
\label{eq:score}
\tag{8}
\end{equation}


The documents are re-ranked based on this score (Algorithm~\ref{alg:mmr-year}), and the top-$K$
 documents by $S(\cdot\mid q)$ are selected as context.

\begin{algorithm}[]
\caption{MMR + year-aware ranking (per query)}\label{alg:mmr-year}
\begin{algorithmic}[1]
\State \textbf{Input:} query $q$, candidates $\mathcal{D}=\{d_1,\ldots,d_n\}$; $n \leq 20$, parameters $\lambda,\alpha$
\For{$i=1$ to $n$}
  \State $s_i\gets \cos\!\big(e(q),e(t(d_i))\big)$
  \State $r_i\gets \max_{j\neq i}\cos\!\big(e(t(d_i)),e(t(d_j))\big)$
  \State $\mathrm{MMR}_i\gets \lambda s_i-(1-\lambda)r_i$
\EndFor
\State compute ${\tau}_i$ for all $i$ using Equation \eqref{eq:year}
\State $S_i\gets \alpha\,\mathrm{MMR}_i+(1-\alpha) {\tau}_i$
\State \textbf{return} re-ranking documents in $\mathcal{D}$ by $S_i$ (descending)
\end{algorithmic}
\end{algorithm}

\subsection{Contradiction Detection Scoring Framework}\label{sec:contradiction}\vspace{-2pt}
To quantify conflicting evidence in the retrieved abstract pool for a query, the subset $\mathcal{R}_{i,j}=\{\,d_1,\ldots,d_n\,\}\subseteq\mathcal{D}_{i,j}$
 was considered. Each document $d\in\mathcal{R}_{i,j}$ is embedded with a scientific encoder $e_{\mathrm{sim}}$ (SPECTER~\cite{cohan2004specter}), producing representation $\mathbf{h}(d) = e_{\mathrm{sim}}(t(d))$. For any abstract pair $(d_a,d_b)\in\mathcal{R}_{i,j}\times\mathcal{R}_{i,j}$, a coarse similarity was computed as:
\[
\operatorname{sim}_{\mathrm{abs}}(d_a,d_b)=\cos\!\big(\mathbf{h}(d_a),\mathbf{h}(d_b)\big),
\tag{9}
\]

\noindent serving as a coarse filter. Each abstract $d$ was segmented into sentences $\mathcal{S}(d)=\{\,s^{(1)}_d,\ldots,s^{(L_d)}_d\,\}$. Sentences were embedded with the same encoder, $\mathbf{h}(s)=e_{\mathrm{sim}}(s)$. For a pair $(d_a,d_b)$, candidate sentence pairs were retained when,
\[
\cos\!\big(\mathbf{h}(s),\mathbf{h}(t)\big)\ \ge\ \theta_{\text{sent}}=0.75,
\qquad (s,t)\in \mathcal{S}(d_a)\times \mathcal{S}(d_b).
\tag{10}
\]
0.75 is selected as the threshold because scores below 0.70 tend to produce false positives, while scores above 0.80 fail to capture subtle medical contradictions.
The resulting candidate set is denoted $\mathcal{P}(d_a,d_b)\subseteq \mathcal{S}(d_a)\times\mathcal{S}(d_b)$. 
Each $(s,t)\in\mathcal{P}(d_p,d_q)$ is passed to a biomedical natural language inference (NLI) classifier $f_{\mathrm{nli}}$ 
(PubMedBERT-MNLI MedNLI~\cite{pritamdeka_pubmedbert_mnli_mednli}) that returns
$(P_{\text{ent}}(s,t),\,P_{\text{neu}}(s,t),\,P_{\text{con}}(s,t))$, 
where $P$ denotes the probability assigned to each relation: 
\emph{entailment} (ent) indicates that the hypothesis is supported by the premise, 
\emph{neutral} (neu) denotes no clear relation, and \emph{contradiction} (con) indicates conflict. 
The contradiction function $\mathcal{CNT}:\mathcal{R}_{i,j}\times\mathcal{R}_{i,j}\to[0,1]$
is then defined at the document level by the peak contradiction probability over candidate sentence pairs:
\[
\mathcal{CNT}(d_a,d_b)=\max_{(s,t)\in\mathcal{P}(d_a,d_b)} P_{\text{con}}(s,t).
\tag{11}
\]

The framework thus reports, for each document pair, i) the document-level similarity, ii) the peak contradiction score, and iii) the most indicative sentence pair with similarity evidence. 
A document-level contradiction salience within the pool was defined as:

\[
\overline{\mathcal{CNT}}(d) \;=\; \frac{1}{|\mathcal{R}_{i,j}|-1}
\sum_{\substack{d'\in\mathcal{R}_{i,j}\\ d'\neq d}}
\mathcal{CNT}(d,d').
\tag{12}
\]

Given $K$, the \emph{most-contradictory} and \emph{least-contradictory} context sets used in the retrieval variants were constructed as:
\[
\mathcal{C}^{\text{most}}_{i,j}=\operatorname*{arg\,topK}_{d\in\mathcal{R}_{i,j}}  \big(\overline{\mathcal{CNT}}(d)\big),
\qquad
\mathcal{C}^{\text{least}}_{i,j}=\operatorname*{arg\,topK}_{d\in\mathcal{R}_{i,j}} \big(-\overline{\mathcal{CNT}}(d)\big),
\tag{13}
\]
respectively, while the \emph{most-similar} condition employed the top-$K$ by the diversity-aware score $S(\cdot)$ in Equation~\eqref{eq:score}. 

\subsection{Retrieval-Augmented Generation Pipeline}\label{sec:rag}
\begin{figure}[t]
  \centering
  \includegraphics[scale=0.90]{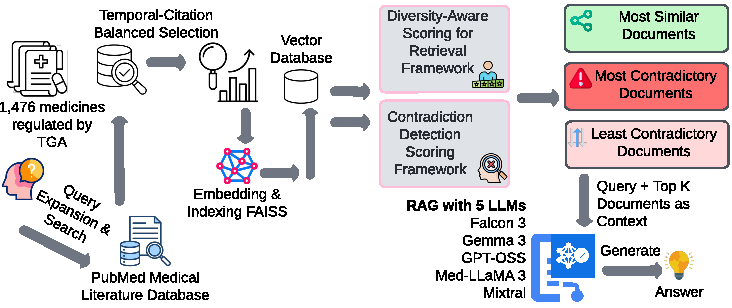}
  \caption{Contradiction-aware medical RAG pipeline, showing data progression from TGA queries through search, embedding, and three retrieval strategies to final LLM-based generation and evaluation.\vspace{-12pt}
}
  \label{fig:rag-pipeline-pdf}
\end{figure}

Given a query $q_{i,j}$ and its ranked list obtained from Equation~\eqref{eq:score}, a context set
$\mathcal{C}_{i,j}=\{d^{(1)},\ldots,d^{(K)}\}$ was formed by taking the top-$K$ documents. A grounded prompting instruction was then applied so that an answer $a_{i,j}^{(\ell)}$ was produced by a model $\ell\in\Lambda$ using only $\mathcal{C}_{i,j}$; if the evidence was insufficient, the token \emph{Insufficient evidence} was to be emitted. 
Three retrieval conditions were instantiated, each with a constant $K$:
i) a \emph{most-similar} condition using the top-$K$ by $S(\cdot\mid q)$;
ii) a \emph{most-contradictory} condition  selecting the $K$ documents with largest contradiction salience (Section~\ref{sec:contradiction}); and
iii) a \emph{least-contradictory} condition  selecting the $K$ smallest by the same criterion. A schematic of the end-to-end pipeline is provided in Figure~\ref{fig:rag-pipeline-pdf}, comprising evidence construction, indexing, and diversity-aware ranking, retrieval variants, RAG inference, and evaluation.

\FloatBarrier
\section{Experimental Setup}\label{sec:setup}
Table~\ref{tab:dataset_stats} summarizes the dataset used in our experiments. 
Starting from nearly 400k retrieved PubMed documents, temporal--citation balanced selection 
retained 91,662 PMIDs, of which 28,873 are unique. Of the 1,476 medicines, 1,074 
have at least one retrieved document, among queries that retained evidence, the mean 
context size is 13.96 documents per query.

\begin{table}[]
\centering
\small
\caption{Statistics of the constructed TGA-PubMed dataset used in our experiments.}
\label{tab:dataset_stats}
\begin{tabular}{l r}
\toprule
\textbf{Statistic} & \textbf{Value} \\[-0.5ex]
\midrule
Medicines (TGA) & 1,476 \\[-0.5ex]
Queries ($6$ per medicine) & 8,856 \\[-0.5ex]
Time span of abstracts & 1975--2025 \\[-0.5ex]
Language & English \\[-0.5ex]
PubMed documents retrieved (raw) & $\sim$400,000 \\[-0.5ex]
Medicines with $\geq$1 retrieved documents & 1,074 \\[-0.5ex]
PMIDs after Temporal-citation balanced selection & 91,662 \\[-0.5ex]
Unique PMIDs in filtered set & 28,873 \\[-0.5ex]
Average documents per query & 13.96 \\[-0.5ex]
\bottomrule
\end{tabular}
\end{table}

The experiments employ the following configurations: MMR parameter $\lambda=0.7$ for balancing relevance and diversity, temporal weighting $\alpha=0.7$ for the final ranking score, and retrieval size $K=5$ documents per query. Document embeddings use BAAI/bge-small-en-v1.5 (384 dimensions), indexed with FAISS HNSW graphs (M=16, ef\_construction=200). PubMed API calls use batch sizes of 300 PMIDs per request. All language models in Table~\ref{tab:models} use temperature=0 for deterministic generation with a 256 token limit. Implementation uses Python 3.10 with LangChain v0.1.0 for RAG orchestration.

Performance assessment employs multiple complementary metrics. Lexical overlap is measured via ROUGE-{1,2,L} ($R_1$, $R_2$, $R_L$)  scores. Semantic similarity uses embedding cosine similarity with the same encoder as retrieval to minimize metric/encoder mismatch. Vector similarity (VSIM), implemented with Gensim, captures term-level relevance through Word2Vec embeddings. Distributional divergence is quantified through Jensen-Shannon divergence (JSD) and Kullback-Leibler divergence (KLD), where lower values indicate closer distributions. All metrics are computed per query and macro-averaged across medicines.

\section{Results and Discussion}\label{sec:results}
\subsection{Overall Performance}
Table~\ref{tab:results-short} presents comprehensive performance metrics across all models and the three retrieval configurations. 
Several key patterns emerge from these results:

\begin{table*}[]
\centering
\small
\setlength{\tabcolsep}{5pt} 
\caption{Macro-averaged results (short references) for 5 LLMs across 3 retrieval conditions. Higher is better except JSD/KLD (↓). Ret.=Retrieval, ms=most similar, mc=most contradicted, lc=least contradicted, cos=cosine, dot=dot-product. VSIM=vector similarity computed with Gensim.}
\label{tab:results-short}\vspace{-6pt}
\begin{tabular}{p{0.65in} l@{\hskip 2pt} c@{\hskip 7pt}c@{\hskip 7pt}c@{\hskip 7pt} c@{\hskip 7pt}c@{\hskip 7pt} c@{\hskip 7pt}c@{\hskip 7pt} c@{\hskip 2pt}c@{\hskip 2pt}}
\toprule
\multirow{2}{*}{\textbf{Model}} & 
\multirow{2}{*}{\textbf{Ret.}} & 
\multicolumn{3}{c}{\textbf{ROUGE}} & 
\multicolumn{2}{c}{\textbf{BERT}} & 
\multicolumn{2}{c}{\textbf{VSIM}} & 
\multirow{2}{*}{\textbf{JSD$\downarrow$}} & 
\multirow{2}{*}{\textbf{KLD$\downarrow$}} \\[-0.5ex]
\cmidrule(lr){3-5} \cmidrule(lr){6-7} \cmidrule(lr){8-9}
& & $R_1$ & $R_2$ & $R_L$ & cos & dot & cos & dot & & \\[-0.5ex]
\midrule
Falcon3-    & ms & 0.154 & 0.027 & 0.103 & 0.589 & 0.589 & 0.708 & 34.07 & 0.213 & 3.130 \\
 7B         & mc & 0.148 & 0.025 & 0.099 & 0.571 & 0.570 & 0.689 & 32.52 & 0.220 & 3.420 \\
              & lc & 0.151 & 0.026 & 0.101 & 0.583 & 0.583 & 0.702 & 33.57 & 0.209 & 3.215 \\[-0.5ex]
\midrule
Gemma-3       & ms & 0.066 & 0.018 & 0.062 & 0.447 & 0.446 & 0.432 & 12.09 & 0.424 & 2.339 \\
              & mc & 0.060 & 0.018 & 0.060 & 0.430 & 0.430 & 0.421 & 11.64 & 0.438 & 2.290 \\
              & lc & 0.064 & 0.017 & 0.060 & 0.437 & 0.437 & 0.428 & 11.98 & 0.427 & 2.373 \\[-0.5ex]
\midrule
GPT-OSS-   & ms & 0.114 & 0.016 & 0.087 & 0.559 & 0.559 & 0.676 & 18.40 & 0.243 & 3.394 \\
    20B & mc & 0.098 & 0.011 & 0.076 & 0.545 & 0.545 & 0.661 & 15.05 & 0.260 & 3.695 \\
      & lc & 0.100 & 0.016 & 0.079 & 0.549 & 0.549 & 0.668 & 16.37 & 0.257 & 3.514 \\[-0.5ex]
\midrule
Med- & ms & 0.156 & 0.032 & 0.109 & 0.573 & 0.573 & 0.645 & 28.69 & 0.256 & 2.772 \\
    LLaMA3-       & mc & 0.134 & 0.027 & 0.097 & 0.529 & 0.529 & 0.607 & 24.75 & 0.283 & 2.988 \\
    8B          & lc & 0.137 & 0.029 & 0.099 & 0.537 & 0.537 & 0.619 & 25.19 & 0.278 & 2.653 \\[-0.5ex]
\midrule
Mixtral-  & ms & 0.163 & 0.029 & 0.118 & 0.601 & 0.601 & 0.706 & 32.74 & 0.214 & 3.208 \\
  8x7B    & mc & 0.131 & 0.022 & 0.096 & 0.551 & 0.551 & 0.670 & 30.33 & 0.225 & 3.752 \\
              & lc & 0.140 & 0.023 & 0.099 & 0.579 & 0.579 & 0.690 & 30.84 & 0.223 & 2.813 \\[-0.5ex]
\bottomrule
\end{tabular}
\end{table*}

\begin{table*}[]
\centering
\small
\setlength{\tabcolsep}{5pt} 
\caption{Macro-averaged results (short references) for 5 LLMs across 3 retrieval conditions. Higher is better except JSD/KLD (↓). Ret.=Retrieval, ms=most similar, mc=most contradicted, lc=least contradicted, cos=cosine, dot=dot-product. VSIM=vector similarity computed with Gensim.}
\label{tab:results-short}\vspace{-6pt}
\begin{tabular}{p{0.65in} l@{\hskip 2pt} c@{\hskip 7pt}c@{\hskip 7pt}c@{\hskip 7pt} c@{\hskip 7pt}c@{\hskip 7pt} c@{\hskip 7pt}c@{\hskip 7pt} c@{\hskip 2pt}c@{\hskip 2pt}}
\toprule
\multirow{2}{*}{\textbf{Model}} & 
\multirow{2}{*}{\textbf{Ret.}} & 
\multicolumn{3}{c}{\textbf{ROUGE}} & 
\multicolumn{2}{c}{\textbf{BERT}} & 
\multicolumn{2}{c}{\textbf{VSIM}} & 
\multirow{2}{*}{\textbf{JSD$\downarrow$}} & 
\multirow{2}{*}{\textbf{KLD$\downarrow$}} \\[-0.5ex]
\cmidrule(lr){3-5} \cmidrule(lr){6-7} \cmidrule(lr){8-9}
& & $R_1$ & $R_2$ & $R_L$ & cos & dot & cos & dot & & \\[-0.5ex]
\midrule
Falcon3-    & ms & 0.154 & 0.027 & 0.103 & 0.589 & 0.589 & \textbf{0.708} & \textbf{34.07} & \textbf{0.213} & 3.130 \\
 7B         & mc & \textbf{0.148} & 0.025 & \textbf{0.099} & \textbf{0.571} & \textbf{0.570} & \textbf{0.689} & \textbf{32.52} & \textbf{0.220} & 3.420 \\
              & lc & \textbf{0.151} & 0.026 & \textbf{0.101} & \textbf{0.583} & \textbf{0.583} & \textbf{0.702} & \textbf{33.57} & \textbf{0.209} & 3.215 \\[-0.5ex]
\midrule
Gemma-3       & ms & 0.066 & 0.018 & 0.062 & 0.447 & 0.446 & 0.432 & 12.09 & 0.424 & \textbf{2.339} \\
              & mc & 0.060 & 0.018 & 0.060 & 0.430 & 0.430 & 0.421 & 11.64 & 0.438 & \textbf{2.290} \\
              & lc & 0.064 & 0.017 & 0.060 & 0.437 & 0.437 & 0.428 & 11.98 & 0.427 & \textbf{2.373} \\[-0.5ex]
\midrule
GPT-OSS-   & ms & 0.114 & 0.016 & 0.087 & 0.559 & 0.559 & 0.676 & 18.40 & 0.243 & 3.394 \\
    20B & mc & 0.098 & 0.011 & 0.076 & 0.545 & 0.545 & 0.661 & 15.05 & 0.260 & 3.695 \\
      & lc & 0.100 & 0.016 & 0.079 & 0.549 & 0.549 & 0.668 & 16.37 & 0.257 & 3.514 \\[-0.5ex]
\midrule
Med- & ms & 0.156 & \textbf{0.032} & 0.109 & 0.573 & 0.573 & 0.645 & 28.69 & 0.256 & 2.772 \\
    LLaMA3-       & mc & 0.134 & \textbf{0.027} & 0.097 & 0.529 & 0.529 & 0.607 & 24.75 & 0.283 & 2.988 \\
    8B          & lc & 0.137 & \textbf{0.029} & 0.099 & 0.537 & 0.537 & 0.619 & 25.19 & 0.278 & 2.653 \\[-0.5ex]
\midrule
Mixtral-  & ms & \textbf{0.163} & 0.029 & \textbf{0.118} & \textbf{0.601} & \textbf{0.601} & 0.706 & 32.74 & 0.214 & 3.208 \\
  8x7B    & mc & 0.131 & 0.022 & 0.096 & 0.551 & 0.551 & 0.670 & 30.33 & 0.225 & 3.752 \\
              & lc & 0.140 & 0.023 & 0.099 & 0.579 & 0.579 & 0.690 & 30.84 & 0.223 & 2.813 \\[-0.5ex]
\bottomrule
\end{tabular}
\end{table*}

\textbf{Model Performance Hierarchy.}
Mixtral consistently achieved the highest performance in the most-similar condition ($R_1$=0.163, $BERT_{cosine}$=0.601), followed closely by Med-LLaMA ($R_1$=0.156, $BERT_{cosine}$=0.573) and Falcon ($R_1$=0.154, $BERT_{cosine}$=0.589). The superior performance of Mixtral can be attributed to its mixture-of-experts architecture, enabling more nuanced processing of medical terminology and context.

\textbf{Impact of Contradictions.}
Across models, performance consistently degraded when moving from most-similar to most-contradictory retrieval conditions. The average $R_1$ score decreased by 18.2\% when models were provided with contradictory documents. This was most pronounced for Mixtral, which showed a 20\% reduction in $R_1$, suggesting that larger models may be more susceptible to conflicting information despite their generally superior performance. Based on the results, performance under the least-contradictory condition was higher than the most-contradictory but still fell short of the most-similar condition.

\textbf{Semantic vs. Lexical Alignment.} 
While ROUGE scores showed substantial variation across conditions, semantic similarity metrics ($BERT_{cosine}$) demonstrated greater stability, with an average decrease of only 8.3\% between most-similar and most-contradictory conditions. This suggests that models maintain conceptual understanding even when struggling with precise lexical generation under contradiction. VSIM showed a similar trend, with stronger alignment under most-similar retrievals and moderate declines under contradictory evidence. JSD values stayed stable across conditions, with only small differences between most-similar and most-contradictory settings. In contrast, KLD showed consistently higher values under contradictory retrievals.

\textbf{Model-Specific Observations.} 
Med-LLaMA, despite being fine-tuned for medical applications, did not consistently outperform general-purpose models. This suggests that domain-specific training may not fully compensate for the challenges posed by contradictory retrieval contexts. GPT-OSS demonstrated the most consistent performance across retrieval conditions, with only a 14\% degradation between best and worst conditions. Its transformer MoE architecture appears to provide robustness against conflicting information, though at the cost of lower peak performance.

\subsection{Contradiction and Diversity-Aware Score Joint Distribution}
To understand how document contradictions interact with the diversity-aware score \(S(\cdot)\), a 2-D frequency table was computed over all documents considered during retrieval. Table~\ref{tab:heatmap} reveals critical insights about the relationship between document diversity-aware score and contradiction likelihood. The highest frequency of documents (5,492) falls in the intersection of high contradiction scores (0.8-1.0) and moderate diversity-aware scores (0.2-0.4). This counterintuitive finding suggests that documents with intermediate topical relevance are most likely to contain conflicting information, potentially due to the same medical concepts but from different temporal or methodological perspectives.

\begin{table*}[]
\centering
\small
\renewcommand{\arraystretch}{1.15}
\setlength{\tabcolsep}{8pt}
\caption{Document frequency by binned diversity-aware similarity score $S$ (columns) and contradiction score $\overline{\mathcal{CNT}}$ (rows). Zero values indicate that no documents fall into those ranges.}
\label{tab:heatmap}
\begin{tabular}{lccccc}
\toprule
\multirow{2}{*}{\textbf{$\overline{\mathcal{CNT}}$}} &
\multicolumn{5}{c}{\textbf{Diversity-Aware Similarity Score $S$}} \\[-0.5ex]
\cmidrule(lr){2-6}
 & $[0,0.2)$ & $[0.2,0.4)$ & $[0.4,0.6)$ & $[0.6,0.8)$ & $[0.8,1]$ \\[-0.5ex]
\midrule
$[0,0.2)$   & \heatcell{695} & \heatcell{1712} & \heatcell{584} & \heatcell{0} & \heatcell{0} \\[-0.5ex]
$[0.2,0.4)$ & \heatcell{816} & \heatcell{2501} & \heatcell{642} & \heatcell{0} & \heatcell{0} \\[-0.5ex]
$[0.4,0.6)$ & \heatcell{1058} & \heatcell{3435} & \heatcell{982} & \heatcell{0} & \heatcell{0} \\[-0.5ex]
$[0.6,0.8)$ & \heatcell{1328} & \heatcell{4983} & \heatcell{1295} & \heatcell{0} & \heatcell{0} \\[-0.5ex]
$[0.8,1]$   & \heatcell{1617} & \heatcell{5492} & \heatcell{1249} & \heatcell{0} & \heatcell{0} \\[-0.5ex]
\bottomrule
\end{tabular}
\end{table*}

\subsection{Temporal Distribution of Contradiction Scores}
\begin{figure}[]
\centering
\includegraphics[width=1\linewidth,height=0.25\textheight,keepaspectratio]
{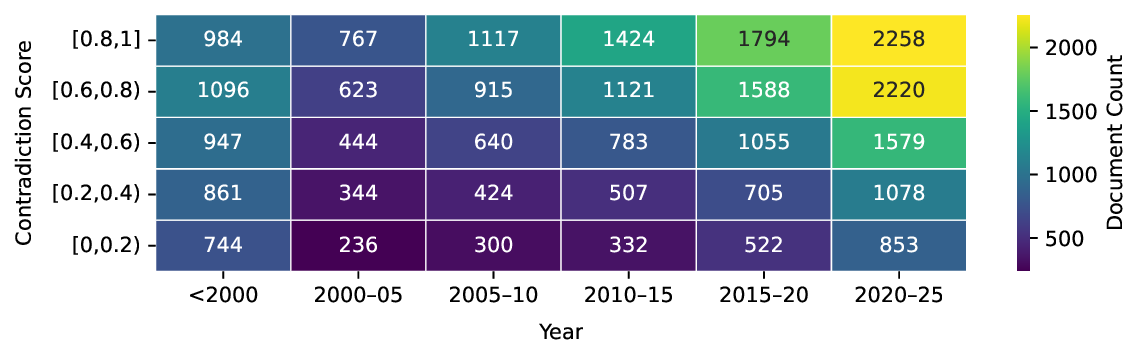}
\vspace{-18pt}\caption{Normalized distribution of documents across contradiction score bins and 5-year publication intervals. }
\label{fig:contradiction_temporal}
\end{figure}

To better understand how contradictions evolve over time, we aggregated documents into 5-year bins and computed proportions across contradiction score bins. Figure~\ref{fig:contradiction_temporal} presents the resulting heatmap.

The results reveal a clear temporal shift in the prevalence of contradictions. 
Before 2000, most documents fall into the lower contradiction score bins, suggesting relatively consistent findings across biomedical literature. 
From the early 2000s onwards, however, the share of documents in higher contradiction bins ([0.6--0.8) and [0.8--1]) rises steadily. By 2010--2025, these bins account for around half or more of all documents per interval, surpassing the lower bins. This indicates that contradictions have become proportionally more prevalent in later years, reflecting both the rapid expansion of biomedical research and the overturning of earlier clinical consensus. These findings underscore the importance of contradiction-aware retrieval strategies that explicitly consider temporal dynamics when integrating evidence into RAG systems.

\vspace{-30pt}
\subsection{Limitations and Future Work}
This study has several limitations. First, our contradiction detection is sentence-level and may miss document-level contradictions. Second, we evaluate only English-language documents, limiting generalization to multilingual medical settings. Third, our model comparisons conflate architecture (MoE vs. dense) with scale (0.27B–46.7B parameters); Mixtral's performance may reflect its size rather than its MoE design. Future work should explore more advanced contradiction-resolution methods, including argumentation mining, evidence synthesis techniques, and integration of clinical guidelines and expert knowledge bases. Additionally, developing specialized medical RAG architectures that model temporal evolution and evidence strength, and comparing models of similar scale to isolate architectural effects, remain important directions for research.

\vspace{-10pt}
\section{Conclusion}\label{sec:conclusion}
This work provides the first systematic evaluation of contradictory information in medical RAG systems through a purpose-built TGA-PubMed dataset across 1,476 medicines. Our analysis shows that contradictions in retrieved evidence consistently degrade model performance, with average $R_1$ scores declining by 18.2\% when contradictory documents are present. Notably, even semantically similar documents often contain conflicting information, with over 5,400 document pairs exhibiting high contradiction scores. These findings show that contradictions in medical literature create a vulnerability for RAG, with a 20\% performance drop unacceptable in high-stakes settings. Future systems must integrate contradiction detection and resolution, using temporal reasoning and uncertainty quantification. Our dataset provides a benchmark for RAG robustness and underscores the need for contradiction-aware architectures to ensure factual accuracy in evolving healthcare contexts.

\begin{table*}[t]
\centering
\small
\setlength{\tabcolsep}{6pt}
\renewcommand{\arraystretch}{1.15}
\caption{Models evaluated in our RAG experiments. We report the *released* model characteristics and the *exact* inference variant used for reproducibility.}
\label{tab:models}
\begin{tabular}{l@{\hskip 6pt} l@{\hskip 6pt} c c l}
\toprule
\textbf{Model (ID)} & \textbf{Family/Architecture} & \textbf{Params} & \textbf{Context} &  \textbf{License} \\
\midrule
Falcon3-7B~\cite{falcon3}        & Decoder-only (dense)     & 7.46B & 32k & TII Falcon \\
gemma-3 (MLX)~\cite{gemma3-270m-it} & Decoder-only (dense)     & 0.27B & 32k & Gemma \\
GPT-OSS-20b~\cite{gpt-oss-20b}          & Transformer MoE          & 21B   & 128k & Apache-2.0 \\
Med-LLaMA3-8B~\cite{med-llama3-8b}      & Llama-3-8B (dense)       & 8.03B & 8k & Llama-3 \\
Mixtral-8x7B~\cite{mixtral-8x7b} & MoE (8×7B; top-2)    & 46.7B & 32k & Apache-2.0 \\
\bottomrule
\end{tabular}
\vspace{3pt}
\end{table*}

\vspace{-6pt}
\section*{Acknowledgments}
This work was initially supported by the Office of the Associate Dean (Research), Faculty of Science, Engineering, and Built Environment, Deakin University, Australia. We also thank Prof. Karin Verspoor and Prof. George Buchanan (RMIT University) and Dr. Diego Mollá-Aliod (Macquarie University) for their valuable intellectual contributions.


\vspace{-6pt}
\section*{Disclosure of Interest}
The authors have no competing interests. This research used publicly available  Consumer Medicine Information data from the Australian Therapeutic Goods Administration, PubMed articles, and open-source LLMs.

\bibliographystyle{splncs04}
\bibliography{x-refs}
\end{document}